\documentclass[%
twocolumn,
 amsmath,amssymb,
 aps, pre
]{revtex4-1}

\usepackage{graphicx}
\usepackage{physics}
\usepackage{siunitx}
\usepackage{booktabs}
\usepackage{subfigure}
\usepackage[OT4]{fontenc}
\usepackage{color}

\newcommand{\red}[1]{#1}
\newcommand{\blue}[1]{#1}

\begin{document}

\preprint{APS/123-QED}

\title{Random sequential adsorption of Platonic and Archimedean solids}

\author{Piotr Kubala}
 \affiliation{M.\ Smoluchowski Institute of Physics, Department of Statistical Physics, Jagiellonian University, \L{}ojasiewicza 11, 30-348 Krak\'ow, Poland.}

\date{\today}

\begin{abstract}
The aim of the study presented here was the analysis of packings generated according to random sequential adsorption protocol consisting of identical Platonic and Archimedean solids. The computer simulations performed showed, that the highest saturated packing fraction $\theta = \num{0.40210(68)}$ is reached by packings built of truncated tetrahedra and the smallest one $\theta = \num{0.35635(67)}$ by packings composed of regular tetrahedra. The propagation of translational and orientational order exhibited microstructural propertied typically seen in RSA packings and the kinetics of 3 dimensional packings growth were again observed not to be strictly connected with the dimension of the configuration space. Moreover, a \red{novel fast overlap criterion for Platonic and Archimedean solids based on separating axis theorem has been described. The criterion, together with other optimizations, allowed to generate} significantly larger packings, which translated directly to a lower statistical error of the results obtained. Additionally, the \red{new} polyhedral order parameters provided can be utilized in other studies regarding particles of polyhedral symmetry.

\end{abstract}

\pacs{Valid PACS appear here}
\maketitle


\section{Introduction}

Packings of objects are a mature field of theoretical, numerical and experimental studies, dating back to ancient times. The problem of effective transportation of cannonballs on ships in colonial era aroused high interest among mathematicians of that time. The first problem of this kind, examined by Thomas Harriot around 1587, was the so-called \textit{cannonball problem} -- how many cannonballs can be arranged in both a square and a pyramid with a square base; in other words: which squares of natural numbers are also pyramidal numbers, \red{which can be reexpressed as diophantic equation $k (k + 1) (2k + 1) / 6 = n^2$ with the smallest solution $k=24$, $n=70$}. \red{Only in 1918 did George Neville Watson prove, that there are no other solutions of this equation \cite{Watson1918}}. Formulating problems regarding packings is surprisingly easy, however solving them tends to be highly complicated. Johannes Kepler was seeking the most optimal way to pack cannonballs. He conjectured, that the densest packing possible of packing density $\pi/\sqrt{18} \approx 0.74$ is achieved by fcc lattice arrangement \cite{Kepler1611}. 200 years later Carl Friedrich Gauss made a step towards a proof by showing, that fcc packing of balls is the most optimal Bravais lattice packing \cite{Gauss1831}. Only in 2017 was the strict, complete proof presented by a mathematical group lead by Thomas Hales that this is the global maximum \cite{Hales2017}.

Nowadays, maximal packings are made use of \red{in} a variety of science fields, from condensed matter physics, where they model \red{crystaline} structures \cite{Kallus2011}, to telecommunication, where \red{they} indicate how to optimize transfer rates \cite{Agrell2009}. \red{Recent works incorporate various shapes, for example Platonic and Archimidean solids \cite{Torquato2009} or two-parameter families of polyhedra \cite{Chen2014}}. Apart from maximal packings, extensive studies of random packings are being conducted, especially regarding \red{the} so-called random close packings (RCP), because their structure resembles one of liquid crystals, \red{amorphous} media, granular matter and various biological systems \cite{Torquato2010}. \red{The term \textit{random} implies the lack of order, while \textit{close packing} means that the neighbouring particles are in contact and continuation of the process which has been used to obtain the packing (e.g. shaking or tapping the containter), no longer increases the packing density. However, as shown by Torquato \cite{Torquato2000}, the definition of RCP cannot be made mathematically precise and} basic properties, such as packing fraction, are highly sensitive to the type of numerical or experimental protocol, which was used to generate them.

\begin{figure*}
	\centering
	\subfigure[]{\includegraphics[width=0.1\linewidth]{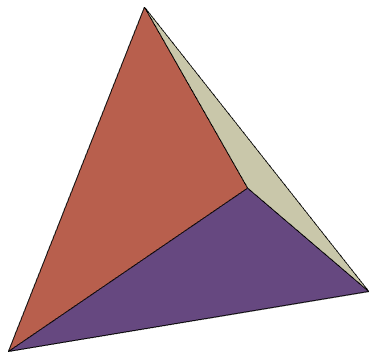}}
	\subfigure[]{\includegraphics[width=0.1\linewidth]{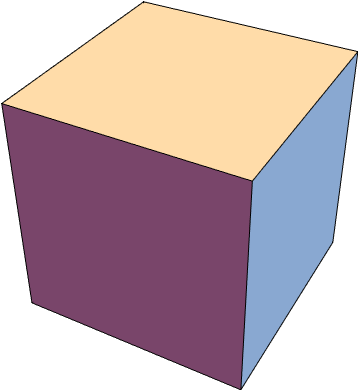}}
	\subfigure[]{\includegraphics[width=0.1\linewidth]{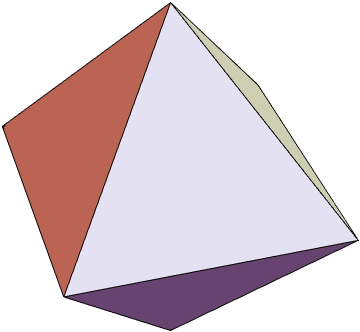}}
	\subfigure[]{\includegraphics[width=0.1\linewidth]{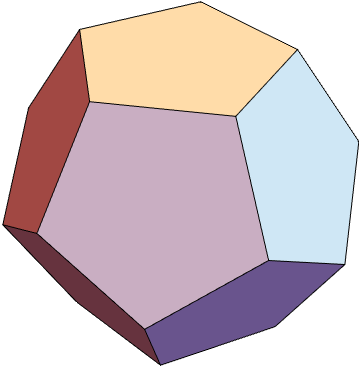}}
	\subfigure[]{\includegraphics[width=0.1\linewidth]{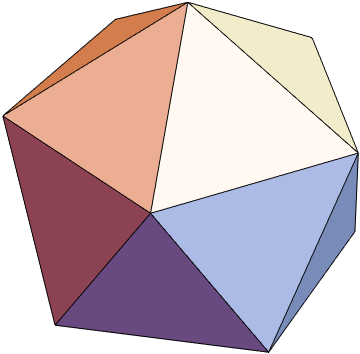}}
	\subfigure[]{\includegraphics[width=0.1\linewidth]{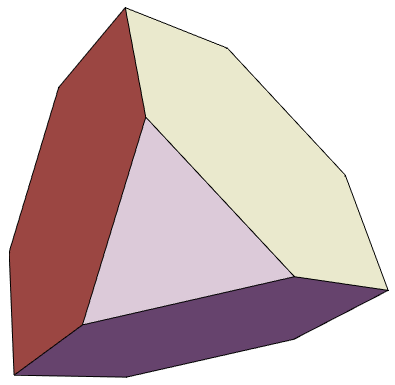}}
	\subfigure[]{\includegraphics[width=0.1\linewidth]{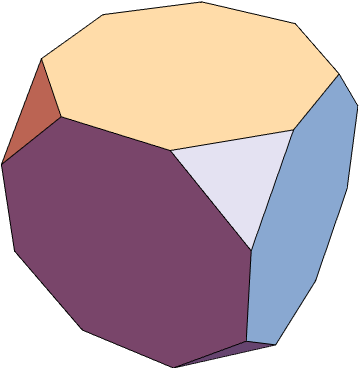}}
	\subfigure[]{\includegraphics[width=0.1\linewidth]{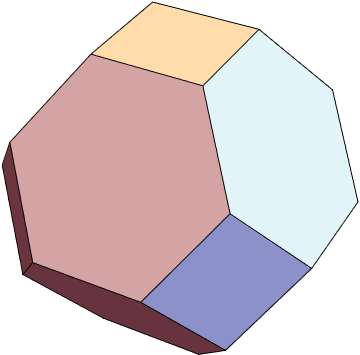}}
	\subfigure[]{\includegraphics[width=0.1\linewidth]{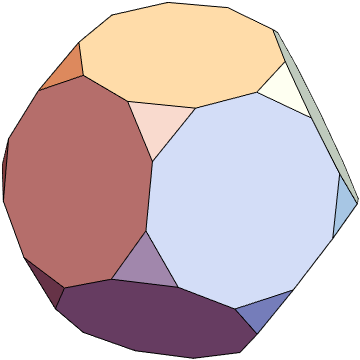}}
	\subfigure[]{\includegraphics[width=0.1\linewidth]{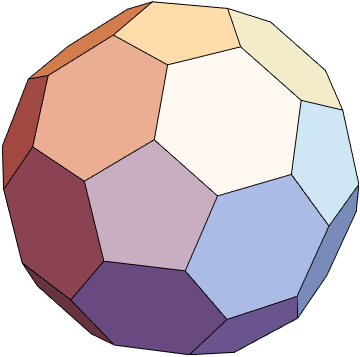}}
	\subfigure[]{\includegraphics[width=0.1\linewidth]{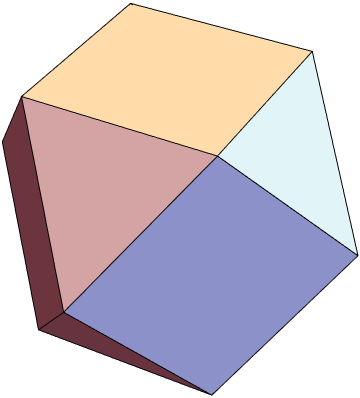}}
	\subfigure[]{\includegraphics[width=0.1\linewidth]{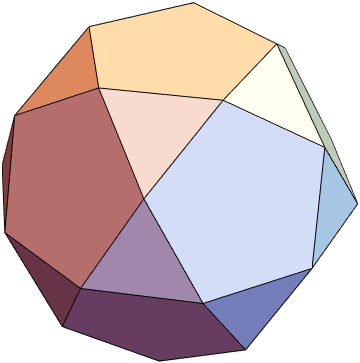}}
	\subfigure[]{\includegraphics[width=0.1\linewidth]{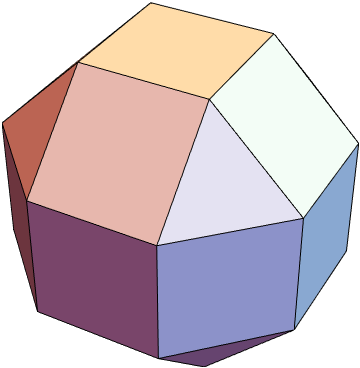}}
	\subfigure[]{\includegraphics[width=0.1\linewidth]{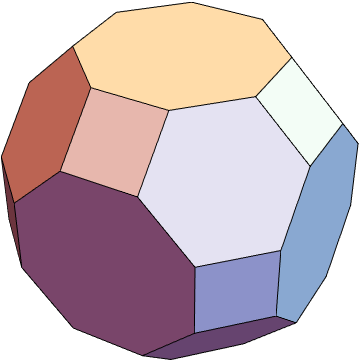}}
	\subfigure[]{\includegraphics[width=0.1\linewidth]{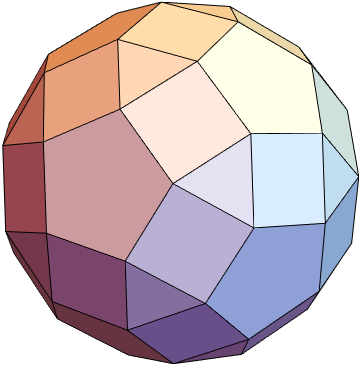}}
	\subfigure[]{\includegraphics[width=0.1\linewidth]{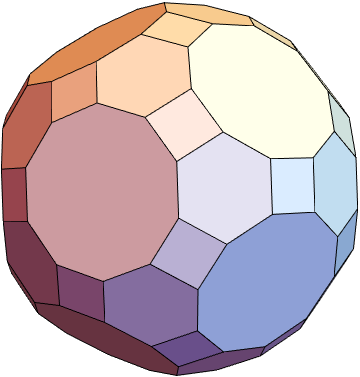}}
	\subfigure[]{\includegraphics[width=0.1\linewidth]{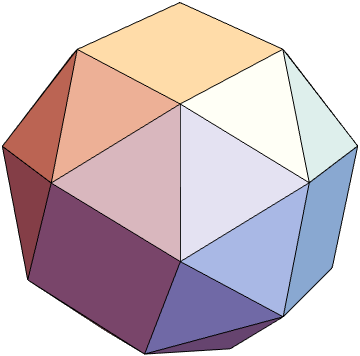}}
	\subfigure[]{\includegraphics[width=0.1\linewidth]{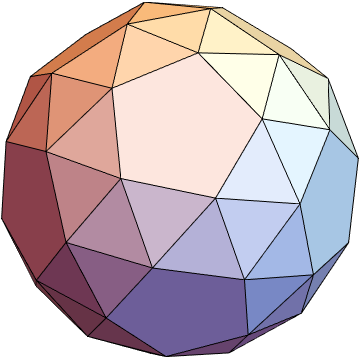}}
	\caption{All Platonic and Archimedean polyhedra. (a) -- regular tetrahedron, (b) -- cube, (c) -- r. octahedron, (d) -- r. dodecahedron, (e) -- r. icosahedron, (f) -- truncated tetrahedron, (g) -- truncated cube, (h) truncated octahedron, (i) -- truncated dodecahedron, (j) -- truncated icosahedron, (k) -- cuboctahedron, (l) -- icosidodecahedron, (m) -- rhombicuboctahedron, (n) -- truncated cuboctahedron, (o) -- rhombicosidodecahedron, (p) -- truncated icosidodecahedron, (q) -- snub cube, (r) -- snub dodecahedron.}
    \label{fig:solids}
\end{figure*}

This study focuses on a slightly different class of random packings, \red{which, contrary to RCP, have well defined mean values}. They are obtained as a result of \red{the} so-called random sequential adsorption (RSA). It is a simple protocol, which consists of subsequent iterations \red{of the following steps:
\begin{itemize}
    \item The position and orientation of a trial object are selected randomly.
    \item If it does not overlap with any of previously placed objects, it is added to the packing and its position and orientation remain unaltered to the end of the process.
    \item If it does overlap, it is removed and abandonded.
\end{itemize}}
\red{The packing is called saturated, when there is no space left for placing any other particles.} Historically, the first person to present the RSA model was Flory, who analyzed the statistics of adjacent pendant groups on long chains of vinyl polymers \cite{Flory1939}. It corresponds to one-dimensional RSA, the \red{mean} saturated packing fraction of which was analytically calculated by Renyi's as a solution to the so-called \textit{car parking problem} \cite{Renyi1958}. RSA is most commonly utilized in two dimensions \cite{Feder1980,Vigil1989,Ciesla2016}, because it models monolayers obtained in irreversible adsorption process \cite{Feder1980}.

3-dimensional RSA packings appear notably more rarely, because there is no physical realization of that process -- it is unclear how a new particle could be placed into the packing, which is already occupied by other particles, and stay in \red{the} place of addition. However, 3-dimensional models are essential to understand RSA process. They can also be used as a toy model for other kinds of packings, because they share common properties with them, for example maximas of density of speroid packings are achieved by similar particle dimensions both for RSA \cite{Sherwood1997} and RCP \cite{Donev2004} packings. There are numerical studies regarding RSA packings of spheres \cite{Zhang2013}, spheroids \cite{Sherwood1997} as well as oriented hypercubes in the context of Palasti conjecture\footnote{\red{$\theta_d=(\theta_1)^d$}, where $\theta_d$ is mean saturated packing density of oriented d-dimensional hypercube \red{RSA} packings, \red{so in particular $\theta_1$ is the packing density in \textit{car packing problem}}.} \cite{Bonnier2001}. Lately, the papers regarding cubes \cite{Ciesla2018Cubes} and cuboids \cite{Kubala2018, Ciesla2018Cuboids} \red{have} been published.

The natural continuation of 3-dimensional RSA studies are Platonic and Archimedean solids (see Fig \ref{fig:solids}). \red{The main goal of this study is to find the mean saturated packing fraction of all 18 polyhedra and to analyze how it is influenced by the solids' shape}. RSA packings of those solids are also interesting in terms of the kinetics of packing growth, \red{which, as it will be presented later, have been lately discovered to be more complex than what had been previously believed}. Moreover, it will also be interesting to compare the result with the recent studies regarding maximal packings \cite{Torquato2009}.

\red{The paper is divided into four section. The second section, Methods, presents the details of computer simulations performed. Subsection A describes their parameters and gives a brief discussion on rotations sampling. The next subsection is devoted to a novel, fast overlap test for Platonic and Archimedean solids which can be generalized to other convex polyhedra. Subsection C indicates where additional optimizations that can be made are described. Subsection A of the next section, Results, present the way to estimate saturated packing fractions and lists their values. Subsection B discusses the kinetics of packing growth, then subsection C elaborates on the correlation between packing fraction and the shape of solids. Subsection D describes the microstructural properties of packings in terms of the propagation of translational and orientational order. It presents new orientational order parameters conforming to point groups of polyhedral symmetries. The last section, Summary, briefly summarizes the findings.}

\section{Methods}

\subsection{Description of simulations}
In order to study the properties of packings of Platonic and Archimedean solids, packings were generated numerically according to RSA scheme \red{described in the Introduction}. For the sake of convenience all solids had a unit volume. The positions of their centers were selected randomly from a cube with the edge size of 50. In order to reduce finite size effects, periodic boundary conditions were used. A brief discussion of errors connected with finite size is presented in section Microstructural properties.

Trial particles, oriented in the same way at the beginning, where rotated randomly in a way which assures that each final orientation is equally probable. A natural way to obtain uniform distribution of $SO(3)$ rotations can be given by a probabilistic measure being translation-invariant Haar measure (as of $SO(3)$ group being compact, right and left Haar measures are equal). It corresponds to the intuition that probability of choosing rotation from a measurable subset $A \subset SO(3)$ should not change after \red{a} translation by an arbitrary rotation $\Gamma$: $P(R \in A)=P(R \in \Gamma \cdot A)$. The details of this reasoning can be found in \cite{Miles1965}. One way of obtaining such a distribution is \red{the} composition of 3 rotations
\begin{equation}
    R = R_3 \cdot R_2 \cdot R_1,
\end{equation}
where $R_1$, $R_2$, $R_3$ are rotations around consecutive coordinate system axes by, respectively, $2\pi x_1$, $\arcsin(2x_2-1)$ and $2\pi x_3$ radians, where $x_1$, $x_2$, $x_3$ are random numbers from $\text{Unif}(0, 1)$ distribution. In \cite{Ciesla2016} there was shown that for such a way of sampling orientations, the correlations of orientations decay to 0 for large enough distances, which is additional, numerical argument for the correctness of that choice.

The main parameter tracked during simulation was the packing fraction defined as
\begin{equation} \label{eq:theta_N}
    \theta(t)=\frac{N(t)V}{V_C}=\frac{N(t)}{V_C},
\end{equation}
where $V=1$ is the particle's volume, $V_C=50^3$ is the system volume and $N(t)$ is the number of particles \red{in the packing} after dimensionless time $t$. Dimensionless time is defined as
\begin{equation}
    t=\frac{n V}{V_C}=\frac{n}{V_C},
\end{equation}
where $n$ is the number of RSA iterations. Dimensionless time is often used to compare the results regardless of $V$ and $V_C$.

The packing becomes saturated, when there is no possibility of adding another object. The mean saturated packing fraction $\theta$ depends only on particle's shape and, contrary to RCP, the mean value is well defined.

For practical reasons, one does not usually generate saturated packings since there is no general method to detect whether a packing is already saturated or not. Hence, in this study packing generation was stopped after \red{arbitrarily chosen time} $t=10^6$ which corresponds to $\num{1.25e11}$ RSA algorithm iterations. \red{This value is a trade-off between the accuracy of estimation of saturated state properties and computational time.}

\red{For each solid 100 independent packings were generated, which gives a few millions of particles in total. As it will be discussed later, it was enough to yield statistical error of mean packing fraction after $t=10^6$ one order lower than the uncertainty introduced by extrapolation to the saturated state.}

\subsection{Overlap detection}

Detecting overlaps of Platonic and Archimedean \red{polyhedra} is the most time consuming operation during RSA packing generation. The choice of the fastest overlap detection algorithm shortens the simulation time significantly and effectively enables obtaining better statistics. In case of polyhedra, commonly used algorithm, especially in computer graphics, is triangulating their surface and testing triangle pairs against intersection. There exist fast triangle-triangle collision tests, such as \cite{Moller1997}. However, there was shown in \cite{Kubala2018} that in case of cuboids another test -- based on separating axis theorem -- can be up to 100 times faster.

Separating axis theorem (SAT) \cite{Gottschalk1996} stands that two convex multidimensional sets are disjunctive if and only if there exists such an axis, that projections of there sets on it are disjunctive. This axis is then called separating axis, hence theorem name. The theorem does not provide any information how to find this separating axis. It turns out, however, that in case of 3D polyhedra, such axis does not exist if none of axes perpendicular to faces of any polyhedron or perpendicular to two edges, each from one polyhedron, is separating axis.

Let $\vb{P}_1,\dots,\vb{P}_n$ be the positions of vertices of polyhedron $P$ and $\vb{u}$ -- normalized vector spanning potential separating axis. The direct way to check whether projections of convex polyhedra $P$ and $Q$ onto $\vb{u}$ overlap is to check the sing of \red{the} expression
\begin{align}
    &\max
    \left\lbrace
        \min_i
        \left\lbrace
            \vb{P}_i\cdot\vb{u}
        \right\rbrace
        ,
        \min_j
        \left\lbrace
            \vb{Q}_j\cdot\vb{u}
        \right\rbrace
    \right\rbrace - \nonumber \\
    &\quad -
    \min
    \left\lbrace
        \max_i
        \left\lbrace
            \vb{P}_i\cdot\vb{u}
        \right\rbrace
        ,
        \max_j
        \left\lbrace
            \vb{Q}_j\cdot\vb{u}
        \right\rbrace
    \right\rbrace.
\end{align}
If it is positive, projections are disjunctive and if it is negative, they overlap. Zero value corresponds to tangent projections. Gottschalk proposed an optimization \red{of this} criterion for cuboids which enables \red{to calculate} all projections in one step \cite{Gottschalk1996}. It can be easily generalized for sufficiently regular polyhedra.

Assume that 8 vertices of a polyhedron form a cuboid or 4 vertices form a rectangle. Let one choose the origin and coordinate system axes $\vu{e}_1$, $\vu{e}_2$, $\vu{e}_3$ so that coordinates of those vertices are $(\pm a, \pm b, \pm c)$, allowing one of $a$, $b$, $c$ to be zero to include \red{the} rectangle case. It is easy to notice, that the half length of a cuboid or rectangle projection onto $\vb{u}$ is then
\begin{align}
    L_\text{C}(\vb{u}) & = a\abs{u^1} + b\abs{u^2} + c\abs{u^3} \nonumber \\ 
    & = a\abs{\vb{u}\vdot\vu{e}_1} + b\abs{\vb{u}\vdot\vu{e}_2} + c\abs{\vb{u}\vdot\vu{e}_3}.
\end{align}
All achiral Platonic and Archimedean solids with octahedral or icosahedral symmetry are built exclusively of concentric, identically oriented groups of vertices arranged in cuboids or rectangles -- it is due to the fact, that both full (achiral) octahedral and icosahedral point groups contain rotations around 3 perpendicular 2-fold axes and reflections through 3 planes spanned by them. It reduces calculation of half length $L_P$ of the polyhedron $P$ to selecting maximal $L_C$. Then, one can test a \red{potential} separating axis for solids $P$ and $Q$ with centres $\vb{O_P}$ and $\vb{O_Q}$ by checking the sign of the expression
\begin{align}
    &\abs{(\vb{O_Q} - \vb{O_P}) \vdot \vb{u}} - \nonumber \\
    &\quad -\qty[\max_{C_i \subset P} \qty{L_{C_i}(\vb{u})} + \max_{C_j \subset Q} \qty{L_{C_j}(\vb{u})}].
\end{align}
Due to this optimization one can significantly reduce the number of calculations performed in an overlap test.

\subsection{Additional optimizations}

\red{Although iterating the RSA steps described in the Introduction without any modifications is enough to obtain RSA packings, several optional optimizations can be made} to increase the speed of packing generation, \red{allowing to obtain better statistic within the same computational time frame. For example, one can utilize the so-called} modified RSA using exclusion zones to reduce the space from which new particles are selected or neighbour lists of adjacent particles to reduce the number of collision tests. \red{To preserve the compact form of the paper these technical detailes have been omitted, however they} can be found in \cite{Zhang2013}. \red{Although the RSA protocol is serial in nature, some parts of the simulation, for example sampling new particles, can be done in parallel. It is discussed in detail} in \cite{Haiduk2018}.

\section{Results}

\subsection{Packing fraction \red{estimation}}

\begin{figure}
	\centering
	\subfigure[]{\includegraphics[width=0.45\linewidth]{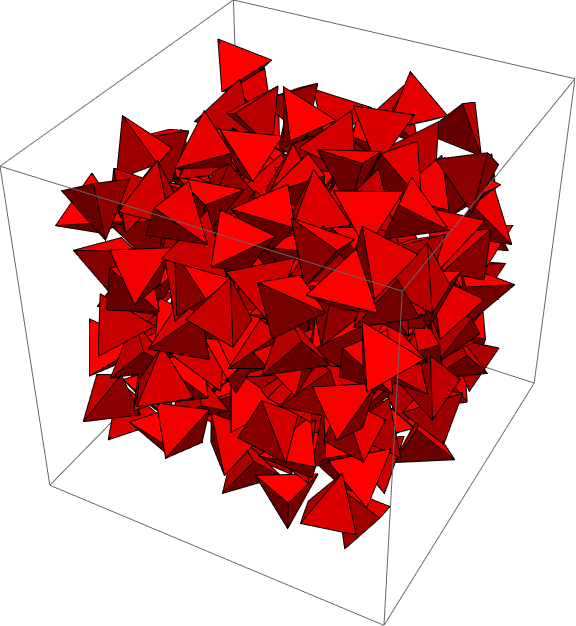}}
	\subfigure[]{\includegraphics[width=0.45\linewidth]{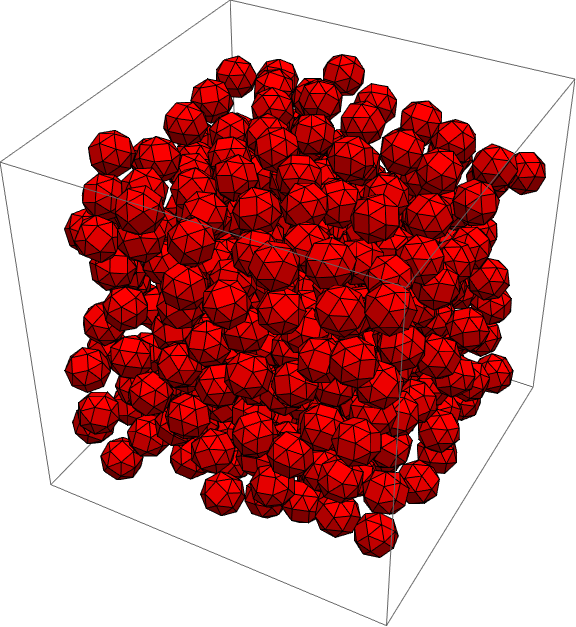}}
	\subfigure[]{\includegraphics[width=0.45\linewidth]{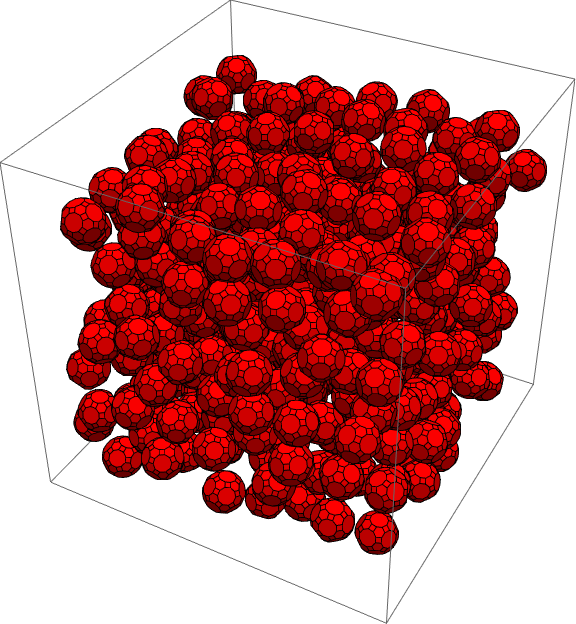}}
	\subfigure[]{\includegraphics[width=0.45\linewidth]{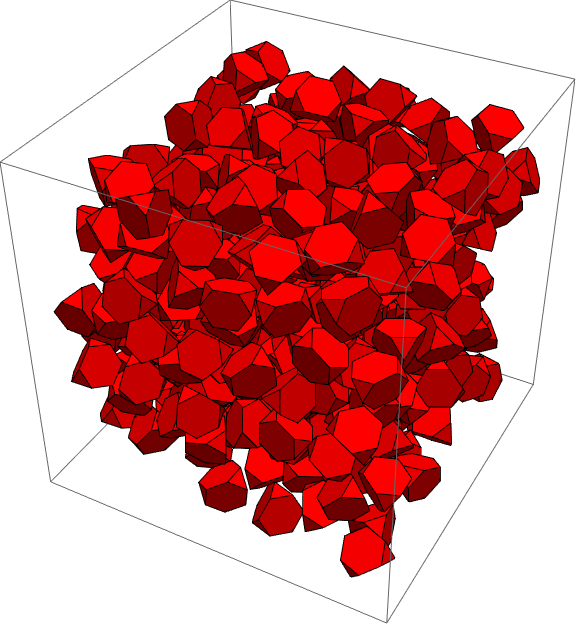}}
	\caption{Example, almost saturated \red{packings} of chosen polyhedra of size $10 \times 10 \times 10$. (a) -- tetrahedron, (b) -- snub cube, (c) -- truncated icosidodecahedron, (d) -- truncated tetrahedron.}
    \label{fig:packings}
\end{figure}

\begin{figure}[htb]
	\centering
    \includegraphics[width=\linewidth]{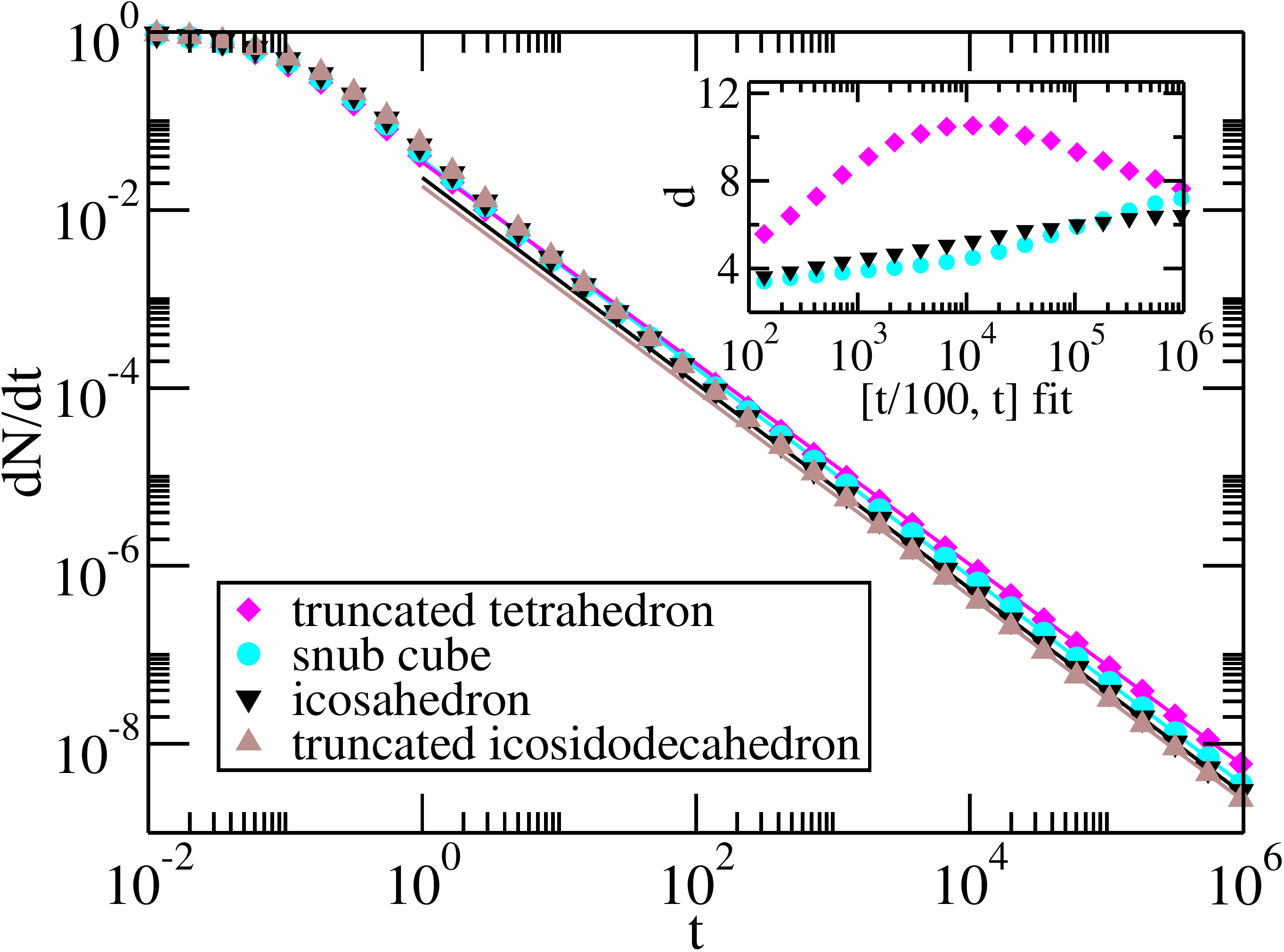}
    \caption{The dependence of $\dv*{N}{t}$ on dimensionless time $t$ (points) with fits to range $[10^4, 10^6]$ (straight lines). Additionally, the inset shows the dependence of the obtained $d$ on time when fitting to range $[t/100, t]$.}
    \label{fig:feder_log}
\end{figure}

Determination of a moment when a packing becomes saturated requires tracking of regions not covered by shapes' excluded volumes. Lately two independent algorithms for 2D shapes -- \cite{Zhang2018} for polygons and \cite{Haiduk2018, Kasperek2018} for ellipses, spherocylinders and rectangles -- have been proposed, however there is no knows generalization in higher dimensions. In case of spheres one can use the Feder's law \cite{Feder1980, Pomeau1980} to extrapolate finite time simulations to infinite time:
\begin{equation} \label{eq:feder}
	\theta - \theta(t) = A t^{-\frac{1}{d}}.
\end{equation}
It is valid for large enough times, where $d$ is the packing dimension and $A$ is a constant. Numerous studies, both analytical \cite{Talbot1989, Baule2017} and numerical \cite{Ciesla2016, Ciesla2018Cuboids} \red{have shown}, that this relation holds for most anisotropic shapes, but with different values of $d$. Eq. (\ref{eq:feder}), after substituting \red{(\ref{eq:theta_N})} and differentiating with respect to $t$ can be rewritten as $\ln(\dv*{N}{t})=\ln(AV_C / d)-(1 / d + 1)\ln(t)$, so $d$ can be determined from linear fit to points $[\ln t, \ln \dv*{N}{t}]$ (see Fig. \ref{fig:feder_log}). Here, the fit was made for $t \in [10^4, 10^6]$. Then, substituting $y=t^{-(1/d)}$ gives $\theta(y) = \theta - Ay$, so $\theta$ is then finally given by intersection of fit to $\theta(y)$ with Y axis.

\begin{table}
	\centering
	\sisetup{
	    table-number-alignment = center
	}
	\begin{tabular}{|l|S[table-format = 1.7(2)]|S[table-format = 1.6(2)]|S[table-format = 1.3(0)]|}
    	\toprule
        \hline
            {Polyhedron name} & {$\theta$} & {$d$} & {$\Psi$} \\
        \hline
        \midrule
    		tr. tetrahedron & 0.40210(68) & 7.631(93) & 0.775 \\
    		tr. cuboctahedron & 0.39631(30) & 6.882(75) & 0.943 \\
    		tr. icosidodecah. & 0.39310(27) & 6.599(87) & 0.970 \\
    		snub cube & 0.39133(30) & 7.235(92) & 0.965 \\ 
    		tr. octahedron & 0.39032(32) & 6.641(71) & 0.910 \\ 
    		cuboctahedron & 0.39032(28) & 6.681(58) & 0.905 \\
    		snub dodecahedron & 0.38978(24) & 6.320(92) & 0.982 \\
    		icosidodecahedron & 0.38922(32) & 6.426(89) & 0.951 \\
    		rhombicuboctah. & 0.38655(32) & 7.080(98) & 0.954 \\
    		rhombicosidodecah. & 0.38640(22) & 6.102(89) & 0.979 \\
    		tr. icosahedron & 0.38610(25) & 6.609(90) & 0.967 \\
    		tr. dodecahedron & 0.38549(22) & 5.930(57) & 0.926 \\
    		icosahedron & 0.38497(30) & 6.443(81) & 0.939 \\
            \textbf{sphere} & 0.3841307(21) & 3.073046(17) & 1.000 \\
    		truncated cube & 0.38075(21) & 5.827(43) & 0.849 \\
    		octahedron & 0.37982(38) & 6.737(70) & 0.846 \\
    		dodecahedron & 0.37936(28) & 6.197(71) & 0.910 \\
    		cube & 0.36249(27) & 6.037(53) & 0.806 \\
    		tetrahedron & 0.35663(67) & 8.119(93) & 0.671 \\
		\hline
	\end{tabular}
    \caption{Extrapolated saturated packing fractions together with corresponding $d$ parameters \red{and sphericity $\Psi$}. The values for sphere are taken from \cite{Zhang2013}. The errors of packing fraction shown are errors propagated from errors of fits. The standard deviation of mean packing fraction after $t=10^6$ was one order of magnitude smaller, so it was not taken into account.}
    \label{tab:results}
\end{table}

The estimated saturated packing fractions are shown in Table \ref{tab:results}. \blue{Moreover, the example packings of size 10x10x10 are shown in Figure \ref{fig:packings} and 3D models of them can be found in Supplemental Material \footnote{See Supplemental Material at [URL placeholder] for STL and Mathematica notebook files containing 3D figures of the packings.}}. The least dense packings are made of regular tetrahedra: $\theta=\num{0.35663(67)}$, and the most dense are made of truncated tetrahedra: $\theta=\num{0.40210(68)}$. To put it in a context, the mean saturated packing fraction of spheres is $\theta=\num{0.3841307(21)}$ placing between them. \red{The errors given are propagated from linear regressions described above. The statistical errors of mean packing fractions before extrapolation have been neglected as they are one order lower.} The new results for cubes differ from the previous ones \cite{Ciesla2018Cubes} on the third decimal digit by the value slightly larger than the tolerance threshold $3\sigma$, both for $\theta$ and $d$. That study utilized smaller simulation time $t=10^5$ which can be the source of the difference.

\subsection{\red{The kinetics of packing growth}}

In papers regarding RSA of cubic shapes \cite{Ciesla2018Cubes,Kubala2018,Ciesla2018Cuboids} one observes the deviation from a typical interpretation of $d$ parameter as the dimension of configuration space. \red{This dimension is equal to 6 in case of all 3D solids without axial symmetry, cuboids, Platonic and Archimidean solids included, while the reported values of $d$ can be as high as 9}. The values of parameter $d$ here, when fitting to $t\in[\num{e4},\num{e6}]$, are for most shapes higher than 6, peaking to \num{8.119(93)} for tetrahedron. For some shapes, namely rhombicosidodecahedron, truncated dodecahedron and cube, they are close to 6 with respect to standard deviation. Apart from having the strong dependence on particle shape, the $d$ values seem to differ when fitting to different ranges of $t$ (see Fig. \ref{fig:feder_log} inset) and it remains unknown whether they ever stabilize. This suggests that the estimated packing fractions can carry systematic errors connected with \red{using slightly shifted $d$ values in extrapolation. Unfortunately, there is no known way of determining their magnitude.} There were attempts to use more sophisticated extrapolation models than the Feder's law, ex. \cite{Ciesla2019}, however they did not render different $\theta$ values, so the simple power fit remains the best approximation. To give the definite values of saturated packing fractions and determine the exact asymptotic behaviour, one needs to develop an algorithm allowing to generate saturated packings in 3 dimensions.

\subsection{\red{The influence of polyhedron shape on packing fraction}}

\begin{figure}
	\centering
    \subfigure[]{\includegraphics[width=0.49\linewidth]{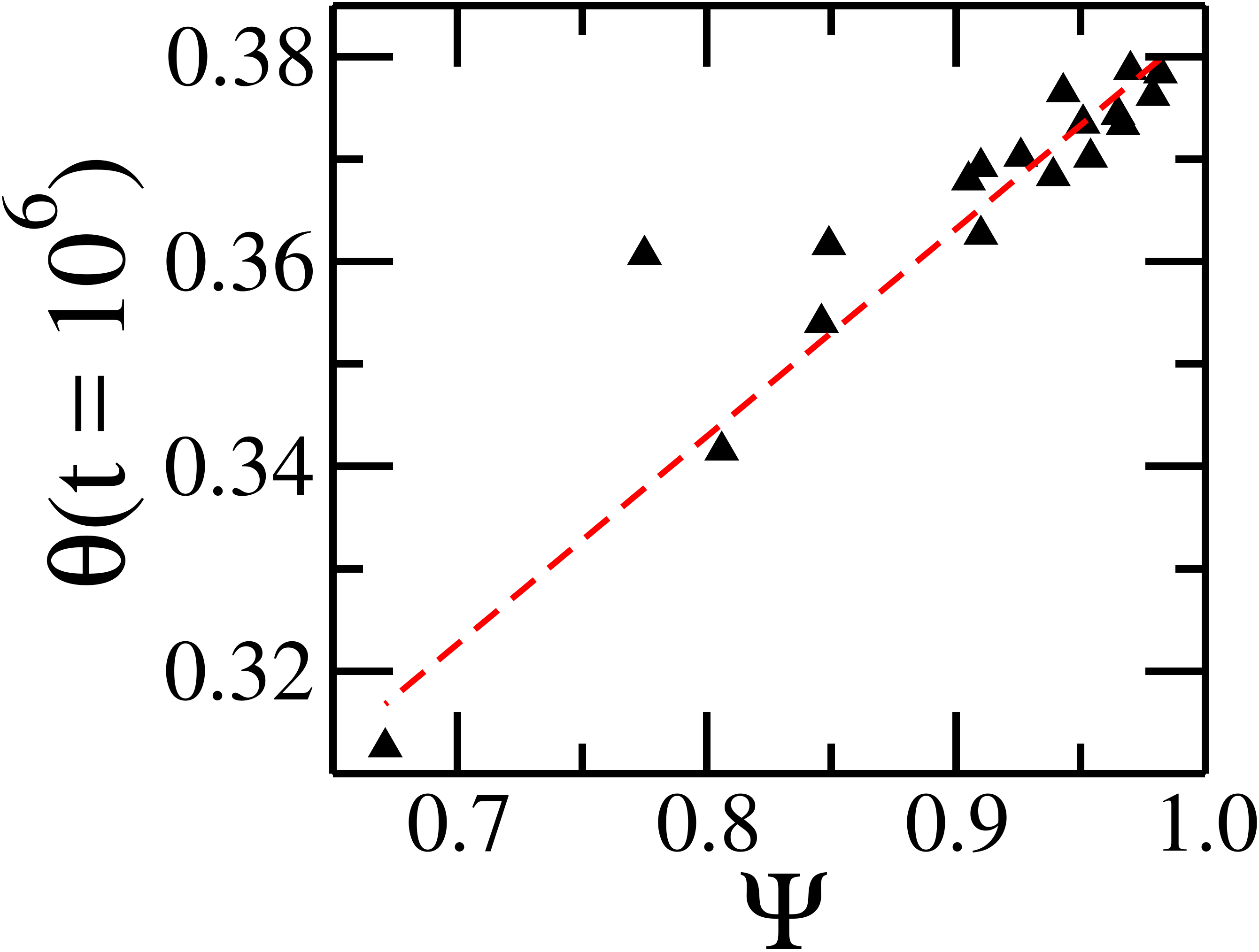}}
	\subfigure[]{\includegraphics[width=0.49\linewidth]{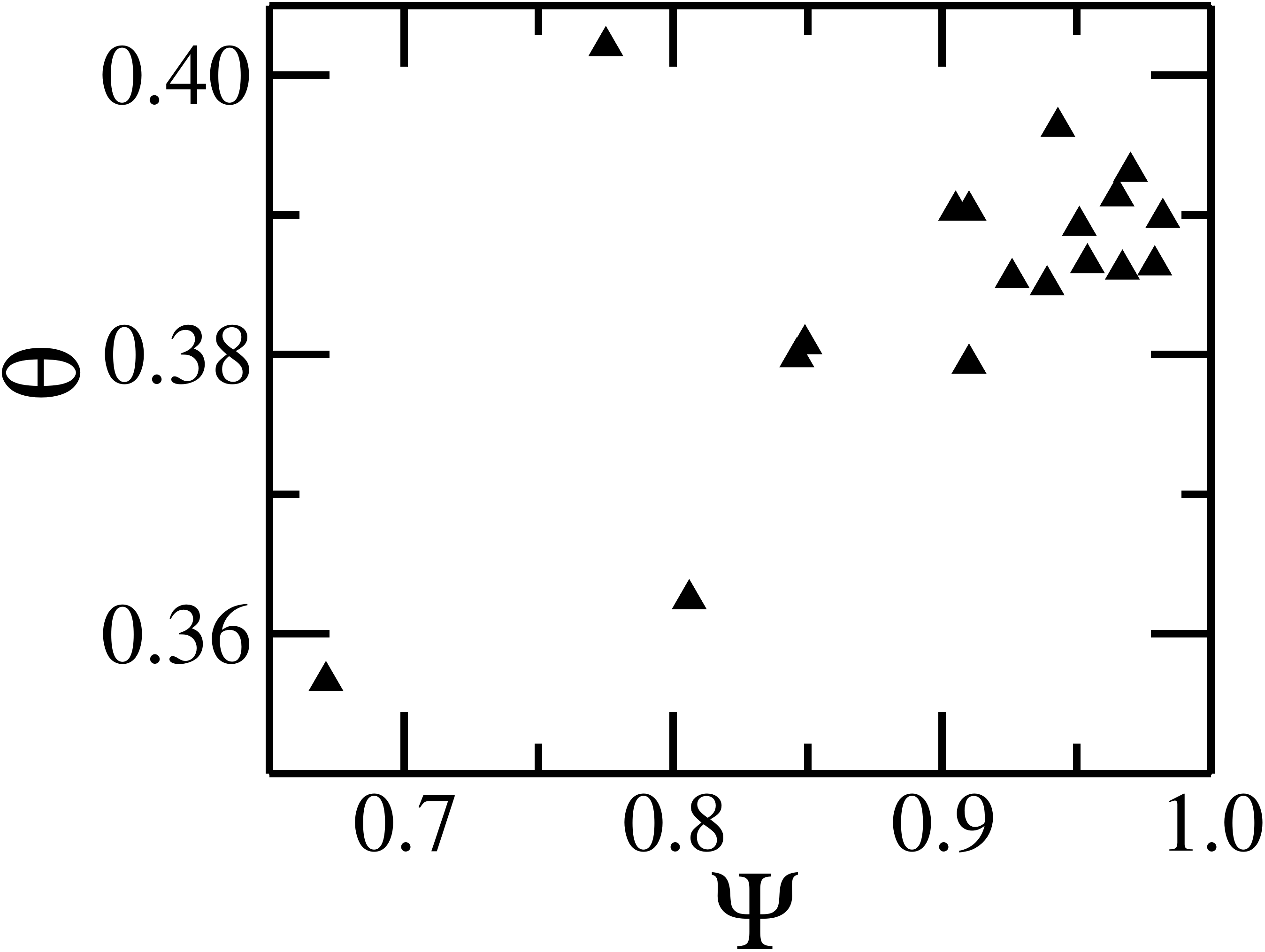}}
    \caption{The dependence of packing fraction $\theta$ on sphericity $\Psi$ of Platonic and Archimedean solids. The left panel shows packing fractions after $t=10^6$, and the right for saturated packings. \red{The dashed line is a linear fit $\theta(t=10^6) = 0.18082 + 0.20263\ \Psi$ to all points excluding truncated tetrahedron.}}
    \label{fig:sphericity}
\end{figure}

Known results for 2D and 3D shapes, such as rectangles \cite{Vigil1989, Kasperek2018}, ellipses, spherocylinders (capsules) \cite{Ciesla2016, Haiduk2018}, dimers \cite{Ciesla2016}, spheroids \cite{Sherwood1997} or cuboids \cite{Kubala2018, Ciesla2018Cuboids} show that the packing fraction grows with the increase of anisotropy in the family of particles of a specific kind until it reaches maximum. In case of studied polyhedra one can see that geometric transformations, such as truncation, rectification and expansion, transforming Platonic solids into Archimedean solids, increase the packing fraction. The most notable difference is for tetrahedron -- after \red{truncating the shape packed}, packings become the most dense from the rarest of all analyzed. One have to notice that those \red{transformations} actually lower the sphericity of polyhedra, defined as \cite{Wadell1935}
\begin{equation}
    \Psi = \frac{\pi^\frac{1}{3}(6V)^\frac{2}{3}}{A},
\end{equation}
which can be used as an indicator of particle anisotropy with volume $V$ and area $A$. $\Psi \in (0,1]$, and maximal value is reached only for sphere. Figure \ref{fig:sphericity} shows the dependence of packing fraction on sphericity of the studied polyhedra. \red{Sphericity values are also included in Table \ref{tab:results}.} The dependence of $\theta(t=10^6)$ on $\Psi$ \red{resembles} linear, \red{with linear fit yielding $R^2 \approx \SI{95}{\percent}$ having excluded truncated tetrahedron}, however, after the extrapolation to $t=\infty$ the differences between points grow. The dependence is neither linear, nor unimodal then -- it shows, that a packing fraction $\theta$ for 3D particles strongly depends on the details of a particle shape. \red{The last statement is supported even more strongly having noted, that the most densely packing shape -- truncated tetrahedron -- lies the farthest from presumptive trends for both time regimes thus its high packing fraction could not be predicted based solely on its sphericity value.}

\red{Densities of} RSA packings of Platonic and Archimedean solids expose significant difference to maximal packings \cite{Torquato2009} \red{in terms of shape dependence}. For that type of packing, Platonic solids have generally higher densities than Archimedean, peaking at 1 for cube, which is different than for RSA packings, where the relation is opposite. On the other hand, tetrahedral shapes show alike behavior -- the least dense maximal packing is for tetrahedron with $\theta = 0.782$ and truncated tetrahedron is one of the most tightly packing shapes with $\theta = 0.958$. Interestingly, the optimal Bravais lattice packing of tetrahedra has the density of $\theta = 0.367$ which is similar to RSA.

\subsection{Microstructural properties}

\begin{figure}
	\centering
	\subfigure[]{\label{fig:corr_not_norm}\includegraphics[width=0.49\linewidth]{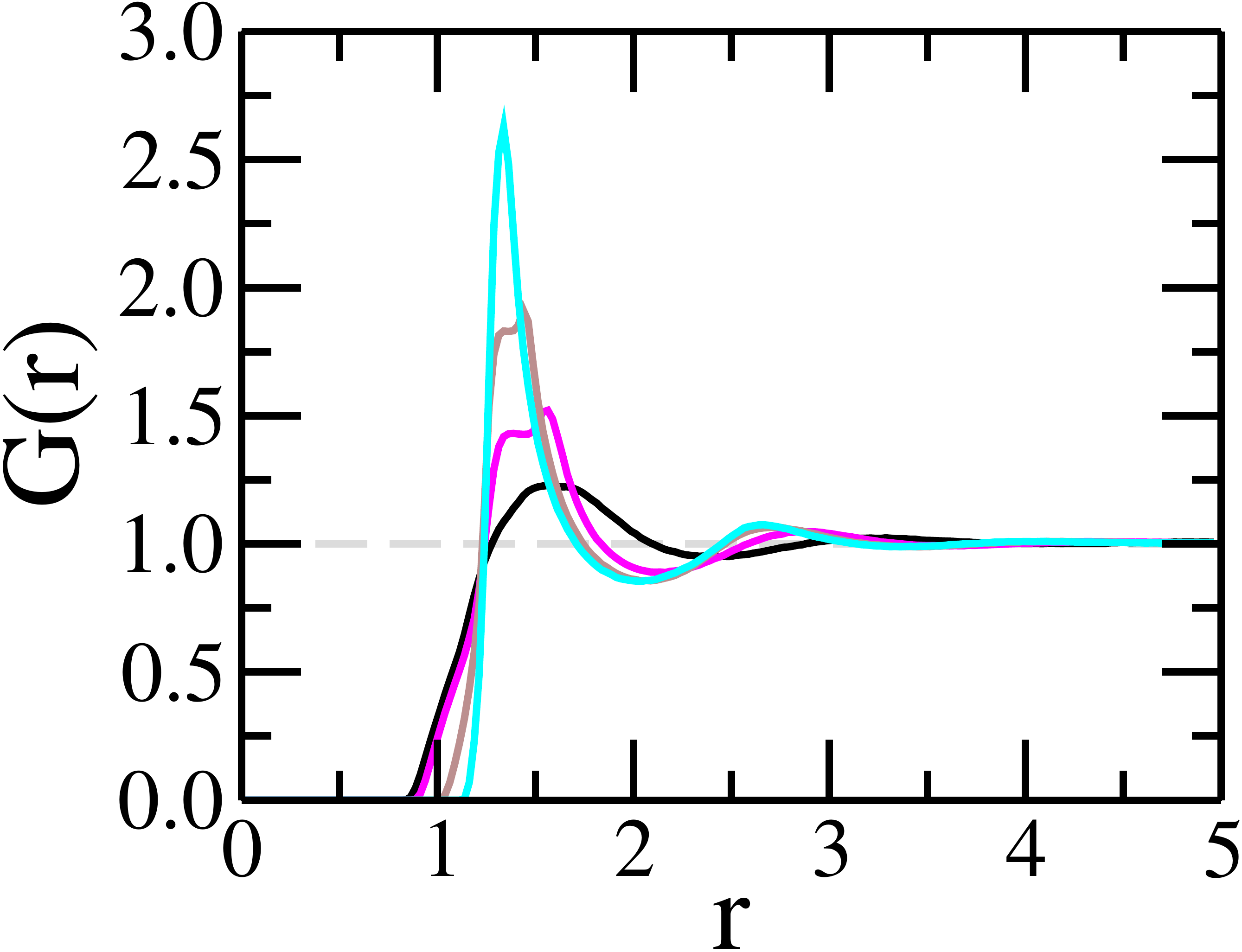}}
	\subfigure[]{\label{fig:corr_norm}\includegraphics[width=0.49\linewidth]{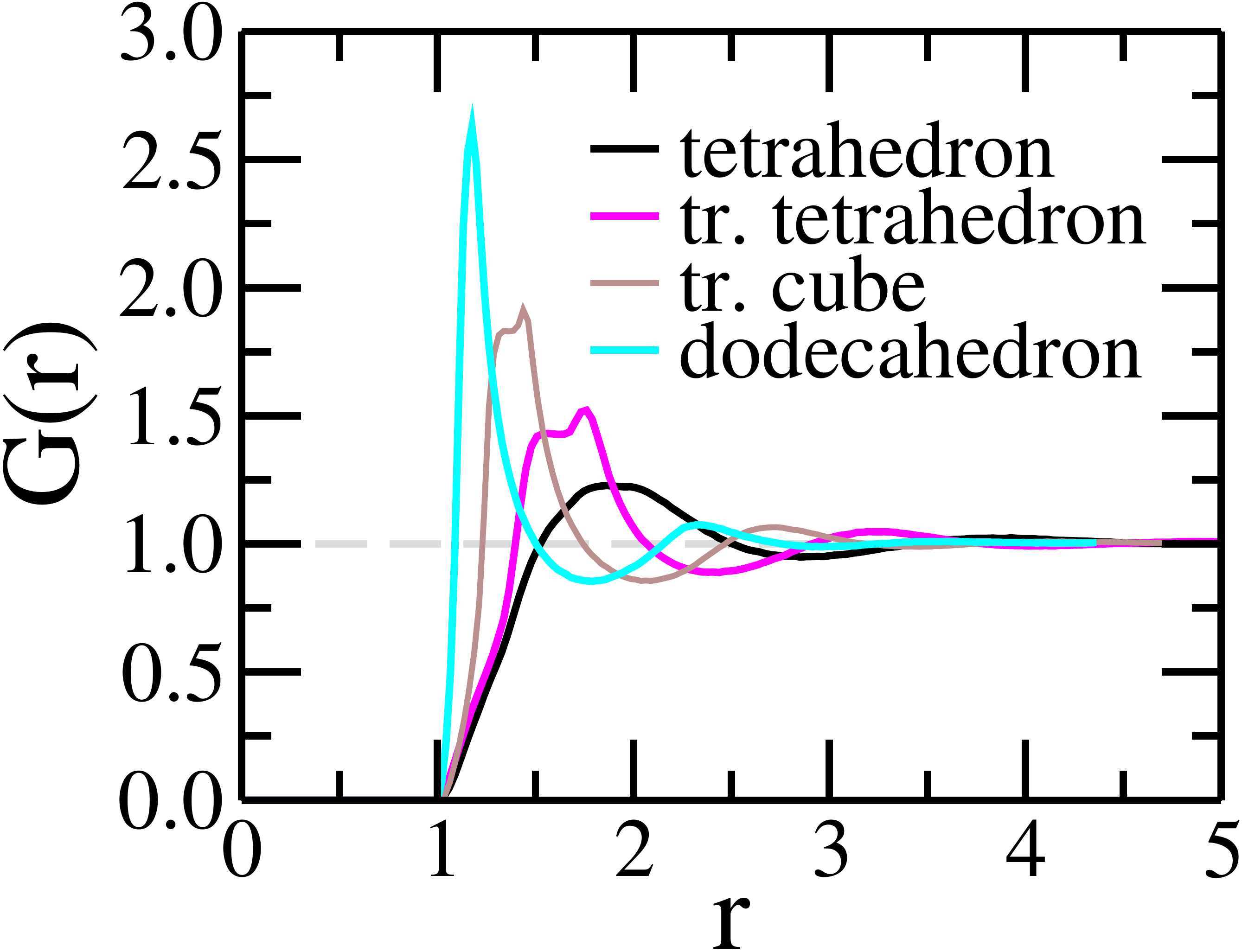}}
	\caption{The dependence of density pair correlation function on the distance between particles. Panel (a) shows real distances in simulations. In panel (b) they are normalized so that the smallest distance possible is equal to 1. \red{The error bars are smaller than the width of the lines.}}
    \label{fig:corr}
\end{figure}

\red{Apart from analyzing global packing parameters, such as packing fraction, one can also investigate into their microstructural properties. Here, they were studied} in terms of density pair correlation function and propagation of orientational order. \red{The latter one required developing new order parameters conforming to polyhedral point symmetries of Platonic and Archimedean solids.}

\subsubsection{\red{Density correlation}}

The density pair correlation function is defined as \cite{Bonnier1994}:
\begin{equation} \label{eq:corr}
    G(r) = \lim_{\dd{r} \to 0} \left< \frac{N(r, r+\dd{r})}{4\pi r^2 \theta \dd{r}} \right>,
\end{equation}
where $N(r, r+\dd{r})$ is the number of pairs whose distance is from $[r, r + \dd{r}]$ interval. Figure \ref{fig:corr} shows $G(r)$ for chosen particle types. Is presents behaviour typically found in RSA packings -- is decays superexponentially \cite{Bonnier1994} with a series of \red{maxima} and \red{minima}. There are additional \red{maxima} for particles with non-equivalent faces, such as truncated tetrahedron. According to \cite{Ciesla2018Boundary}, $G(r)$ is strictly connected with an error in packing fraction introduced by finite size effects, so, as there are almost no correlations after $r=5$, finite size effects should be negligible for a packing size used in this study.

\subsubsection{\red{Orientational order parameters}}

In order to measure full orientational order propagation, the order parameters conforming to polyhedral groups of point symmetries were used:
\begin{subequations} \label{eq:polyhedral}
    \begin{align}
        \rho_4(r) &= \lim_{\dd{r} \to 0} \frac{9}{32}
        \left<
            \sum_{i,j}
            \qty(
                \vb{u_i} \cdot \vb{v_j}
            )^3
        \right>_{[r, r+\dd{r}]}, \label{eq:tetrahedral} \\
        \rho_8(r) &= \lim_{\dd{r} \to 0} \frac{1}{6}
        \left<
            5 \qty[
                \sum_{i,j}\qty(
                    \vb{u_i} \cdot \vb{v_j}
                )^4
            ] - 9
        \right>_{[r, r+\dd{r}]}, \label{eq:octahedral} \\
        \rho_{20}(r) &= \lim_{\dd{r} \to 0} \frac{25}{192} \times \nonumber \\
        & \qquad \times \left<
            7 \qty[
                \sum_{i,j}\qty(
                    \vb{u_i} \cdot \vb{v_j}
                )^6
            ] - 36
        \right>_{[r, r+\dd{r}]}, \label{eq:dodecahedral}
    \end{align}
\end{subequations}
where $\rho_4$, $\rho_8$ and $\rho_{20}$ are for, respectively, tetrahedral, octahedral and icosahedral point groups. $\vb{u_i}$, $\vb{v_j}$ are the smallest sets of normalized equivalent rotational symmetry axes -- namely 3- 4- and 5-fold axes respectively -- for two particles. The summation goes over all pairs of axes, each from one particle, and the average is calculated for all pairs of particles whose centres' distance is in $[r,r+\dd{r}]$ interval. The exponents are even for symmetries where orientation of a particle is fully determined only by orientations of considered axes and odd where both orientation and sense of axes is needed. The values of the exponents are the smallest for which the sum is not constant regardless of particles' orientations. The appropriate linear normalization ensures that the value of these order parameters is 0 for isotropic ensemble of particles and 1 for identically oriented particles. For $\rho_4$ only, the normalization depends on the individual choices of sense of each of 4 axes -- here it has been assumed, that the ends of axes form regular tetrahedron when the beginnings are in the same point. It is also worth noting that $\rho_4$, $\rho_8$, \red{$\rho_{20}$} parameters are suitable for point groups both with and without reflections, namely for both chiral and achiral particles. 

One can also check nematic order, using the standard $P_1$ and $P_2$ parameters
\begin{equation} \label{eq:nematic_abs}
    \rho_n(r) = \lim_{\dd{r} \to 0}\left<P_m \qty(\max_{i,j}\qty{\abs{\vb{u_i} \cdot \vb{v_j}}})\right>_{[r, r + \dd{r}]},
\end{equation}
however, in order to take degenerate axes into account, one have to check all pairs and choose the dot product with the maximal absolute value. The side effect of this operation is that $\rho_n$ is not zero even for isotropic set. $P_2$ is used when both senses of axis are equivalent, $P_1$ when not. Nematic order parameters are suitable for Platonic solids, because they have one class of characteristic axes, which go through the middle of the faces. For Archimedean solids the choice is ambiguous.

\begin{figure}
	\centering
	\subfigure[]{\includegraphics[width=0.49\linewidth]{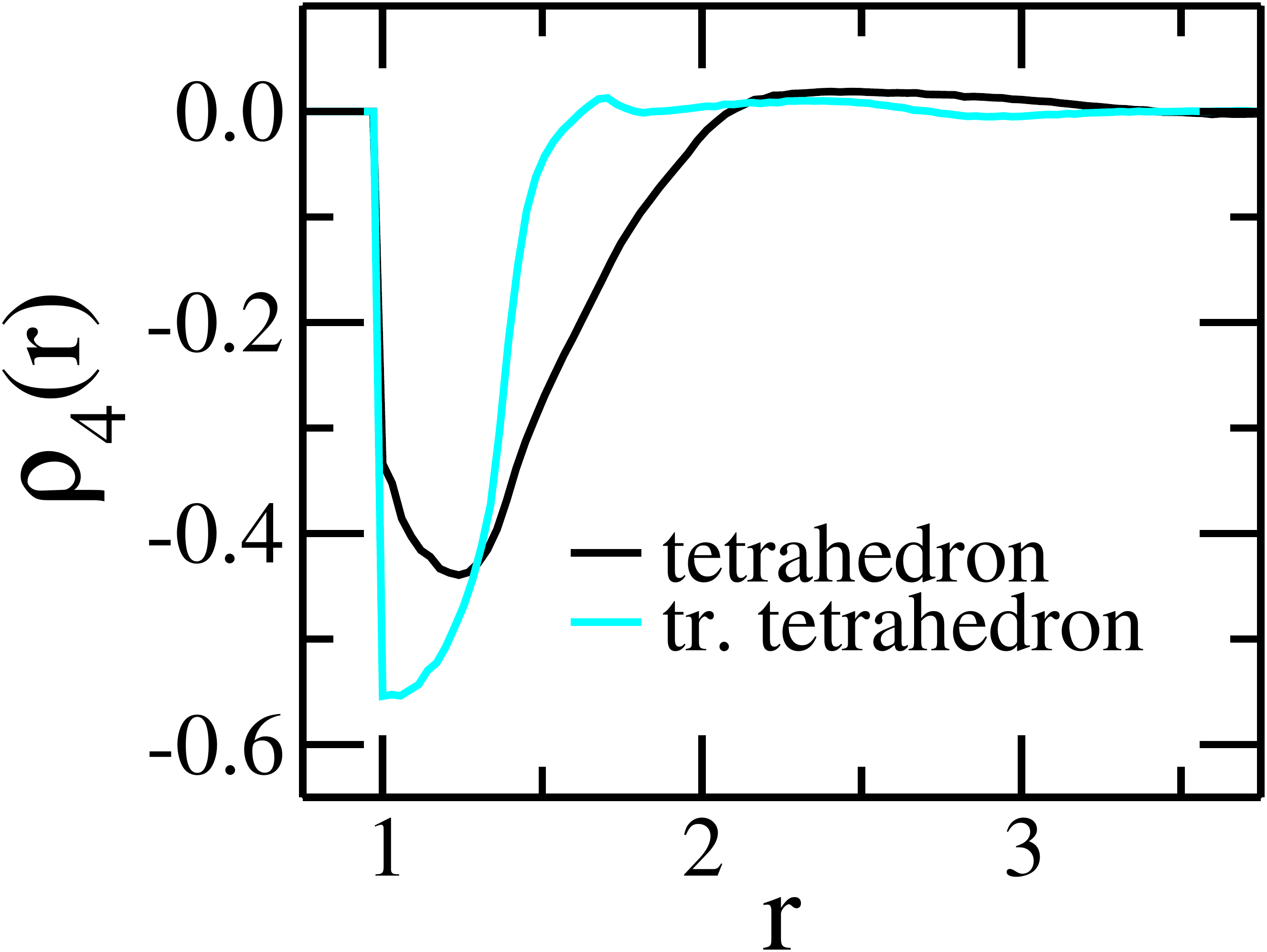}}
	\subfigure[]{\includegraphics[width=0.49\linewidth]{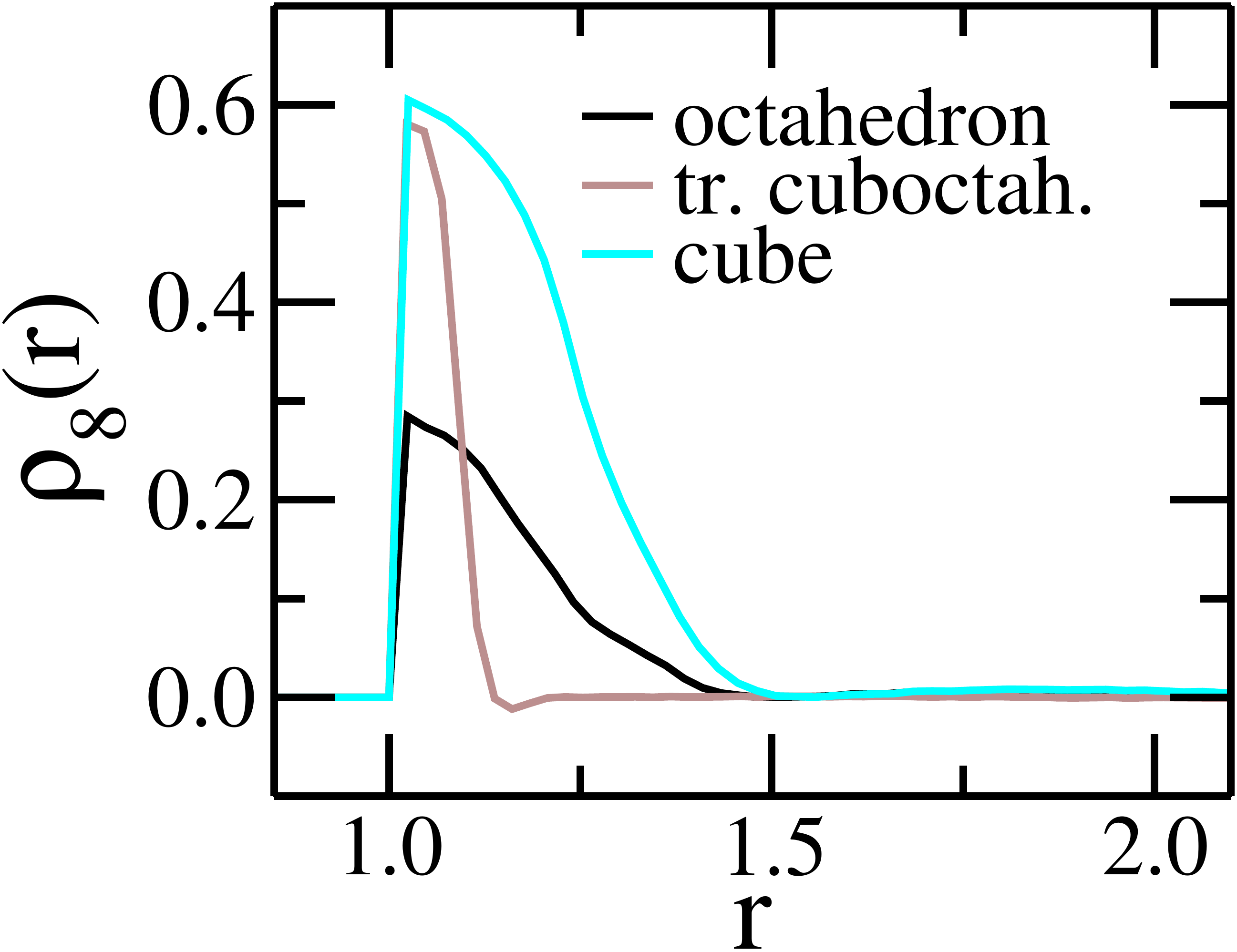}}
	\subfigure[]{\includegraphics[width=0.49\linewidth]{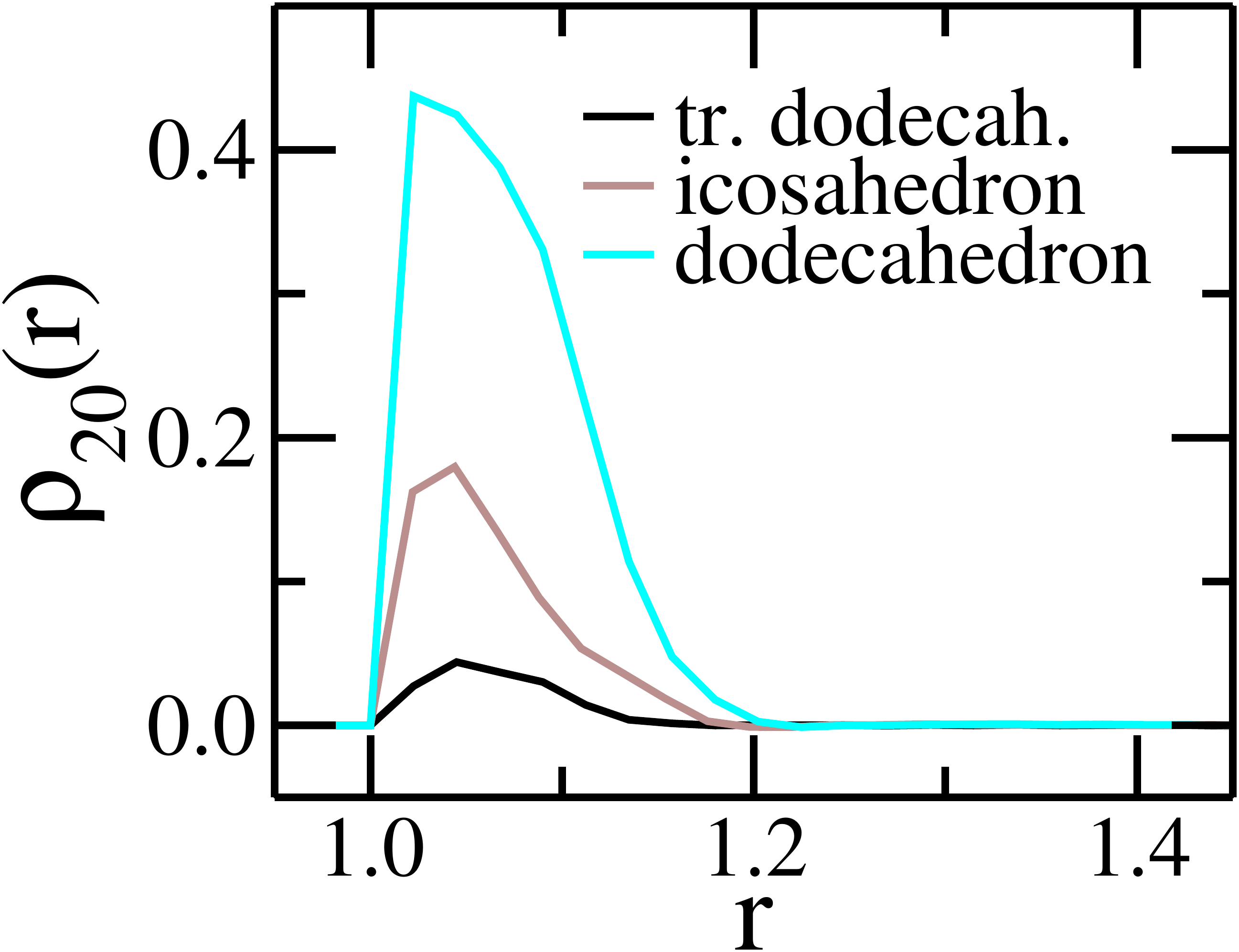}}
	\subfigure[]{\label{fig:order_n}\includegraphics[width=0.49\linewidth]{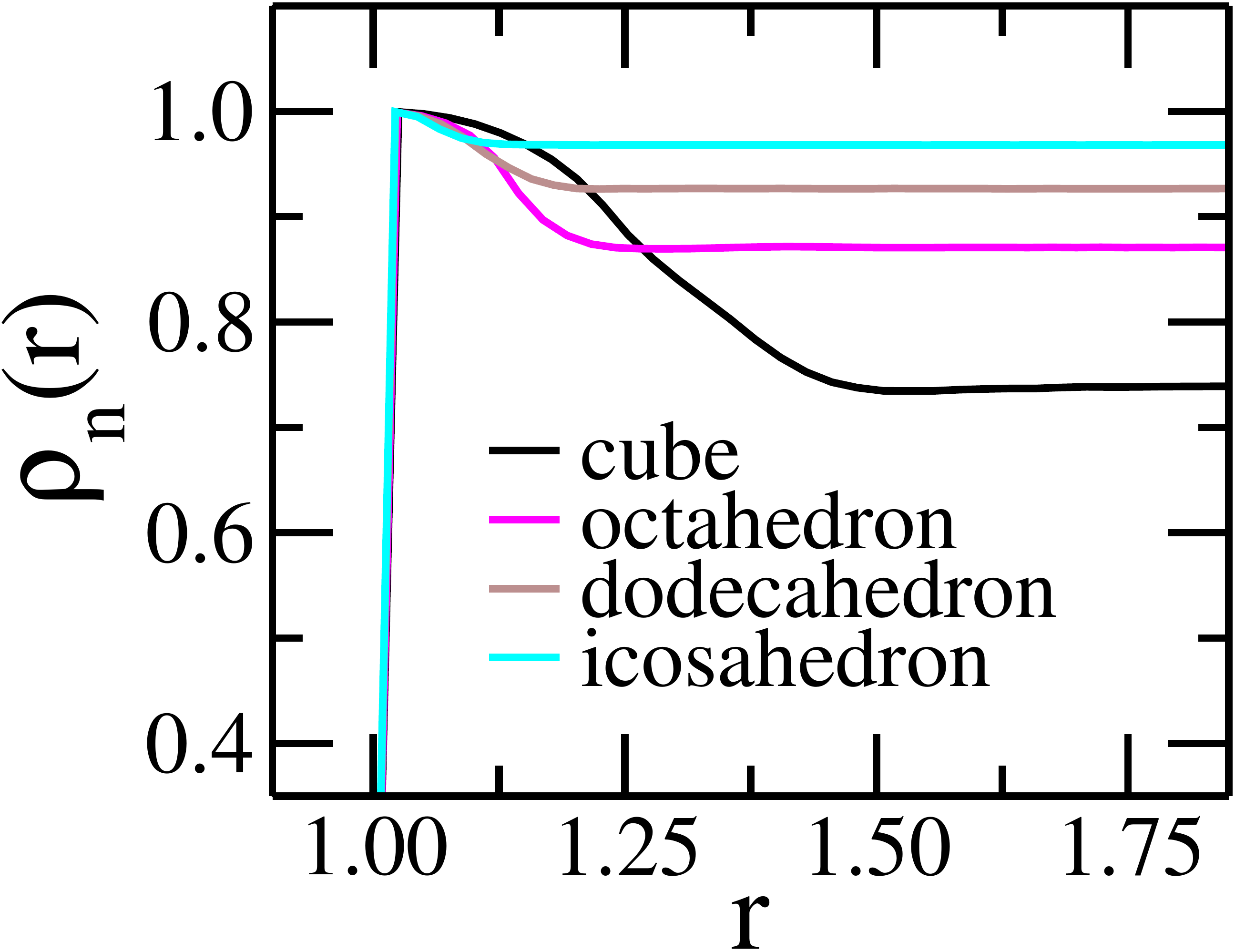}}
	\caption{(a)-(c) The dependence of \blue{polyhedral order $\rho_x$} for chosen Platonic and Archimedean solids on the distance between their centres. The X axis is normalized in such a way that the smallest possible distance is 1. Additionally, panel (d) shows nematic order $\rho_n(r)$.}
    \label{fig:order}
\end{figure}

The dependence of full order parameters on distance is shown in Fig. \ref{fig:order} (a)-(c). Is is typical for RSA packings. The highest order is seen for almost touching particles, where their faces have to be aligned to prevent an overlap, however full order is never achieved. It is due to the fact, that aligned particles still have rotational freedom around the normal axis of close faces. Nematic order parameters (d) confirm full nematic order for small distances. Both full and nematic order parameters decay quickly with a distance, which is also typical for RSA packings \cite{Ciesla2013}.

\section{Summary}

Within the study RSA packings of 5 Platonic and 13 Archimedean solids where examined. It has been shown that the loosest packings are formed by tetrahedra with packing fraction $\theta=0.35663(67)$ and the densest are made of truncated tetrahedra with packing density of $\theta=0.40210(68)$. In general, Archimedean solids form denser packings than Platonic solids, however exact $\theta$ values depend strongly on particle shape, not only on sphericity $\Psi$. For majority of the polyhedra studied the exponent $d$ describing packing growth kinetics is not equal to configuration space dimension, however one needs to generate strictly saturated packings to give the definite answer. There was no global translational or orientational order observed. Additionally, rapid intersection tests and order parameters conforming to polyhedral symmetries have been developed.

\section*{Acknowledgments}

This work was supported by grant no. 2016/23/B/ST3/01145 of the National Science Center, Poland. Numerical simulations were carried out with the support of the Interdisciplinary Center for Mathematical and  Computational Modeling (ICM) at the University of Warsaw under grant no. G-27-8.

\bibliography{main}

\end{document}